\documentclass[aps,twocolumn ,nofootinbib,superscriptaddress,geometry,preprintnumbers]{revtex4-1}

\usepackage{amsmath,amssymb} 
\usepackage{accents}
\usepackage{nccmath}
\usepackage{color}
\usepackage[bookmarks=true,hyperfigures=true]{hyperref}
\usepackage{graphicx}
\usepackage{tensor}
\usepackage{empheq}
\usepackage{nicefrac}
\usepackage{shuffle}
\usepackage{slashed}

\usepackage{feynmp-auto}
\usepackage{feynmp} 
\allowdisplaybreaks[3]

\usepackage[font=small,labelfont=bf,width=0.85\textwidth]{caption}

\let\oldbfseries=\bfseries
\let\oldmdseries=\mdseries
\let\oldnormalfont=\normalfont
\renewcommand{\bfseries}{\oldbfseries\boldmath}
\renewcommand{\mdseries}{\oldmdseries\unboldmath}
\renewcommand{\normalfont}{\oldnormalfont\unboldmath}

\makeatletter
\newlength{\apb@width}
\newcommand{\autoparbox}[2][c]{\settowidth{\apb@width}{#2}\parbox[#1]{\apb@width}{#2}}

\makeatother

\newcommand{\nn}{\nonumber}

\newcommand{\ttau}{\tilde\tau}

\newcommand{\remark}[2][.]{{\color{red}\renewcommand{\bfdefault}{b}\rmfamily\if.#1\else\textbf{#1:} \fi#2}}
\newcommand{\red}{
}

\newcommand{\ee}{\end{equation}}
\newcommand{\beq}{\begin{equation}}
\newcommand{\eeq}{\end{equation}}

\newcommand{\bma}{\begin{pmatrix}}
\newcommand{\ema}{\end{pmatrix}}
\newcommand{\ba}{\begin{eqnarray}}
\newcommand{\ea}{\end{eqnarray}}
 
\newcommand{\vct}{\textbf}
\newcommand{\Order}{\mathcal{O}}

\ifx\genfrac\sdflkaj\else\fi

\newcommand{\ft}[2]{{\textstyle\frac{#1}{#2}}}

\newcommand{\cO}{\mathcal{O}}

\def\l<{\langle}\def\r>{\rangle}

\makeatletter
\newcommand{\namedref}[2]{\hyperref[#2]{#1~\ref*{#2}}}
\newcommand{\secref}{\@ifstar{\namedref{Section}}{\namedref{sec.}}}
\newcommand{\subsecref}{\@ifstar{\namedref{Subsection}}{\namedref{subsec.}}}
\newcommand{\appref}{\@ifstar{\namedref{Appendix}}{\namedref{app.}}}
\newcommand{\tabref}{\@ifstar{\namedref{Table}}{\namedref{tab.}}}
\newcommand{\figref}{\@ifstar{\namedref{Figure}}{\namedref{fig.}}}
\makeatother

\def\[{\begin{equation}}
\def\]{\end{equation}}
\def\<{\begin{eqnarray}}
\def\>{\end{eqnarray}}

\newcommand{\eqn}[1]{(\ref{#1})}

\def\bea{\begin{align}}
\def\eea{\end{align}}
\def\be{\begin{equation}}
\def\ee{\end{equation}}

\begin{document}
\newtheorem{theorem}{Theorem}
\newtheorem{theorem1}{Theorem}
\thispagestyle{empty}


\title{Breakdown of the classical double copy for the effective action \\ of dilaton-gravity 
at NNLO}
\author{Jan Plefka}
\affiliation{Institut f\"ur Physik und IRIS Adlershof, Humboldt-Universit\"at zu Berlin, 
Zum Gro{\ss}en Windkanal 6, 12489 Berlin, Germany}
\author{Canxin Shi}
\affiliation{Institut f\"ur Physik und IRIS Adlershof, Humboldt-Universit\"at zu Berlin, 
Zum Gro{\ss}en Windkanal 6, 12489 Berlin, Germany}
\author{Jan Steinhoff}
\affiliation{Max-Planck-Institut f\"ur Gravitationsphysik
(Albert-Einstein-Institut)\\
M\"uhlenberg 1, D-14476 Potsdam, Germany}
\author{Tianheng Wang}
\affiliation{Institut f\"ur Physik und IRIS Adlershof, Humboldt-Universit\"at zu Berlin, 
Zum Gro{\ss}en Windkanal 6, 12489 Berlin, Germany}
\email{jan.plefka@physik.hu-berlin.de}
\preprint{HU-EP-19/14, SAGEX-19-11-E}

\begin{abstract}
We demonstrate that a recently proposed classical double copy procedure to construct the effective action of two massive particles in dilaton-gravity from the analogous problem of two color charged particles in Yang-Mills gauge theory fails at the next-to-next-to leading order perturbative expansions, i.e. at the 3rd order in the post-Minkowskian and the 2nd order in the post-Newtonian expansion.
\end{abstract}

\maketitle

\section{Introduction and Conclusions}

A spectacular success in fundamental physics has been the detection of gravitational waves
at the LIGO/Virgo detectors since 2015 \cite{Abbott:2016blz}. This detection and analysis
hinges crucially on waveform templates emerging
from high-precision theoretical predictions  in general relativity including higher order perturbative
computations of the two body effective potential in the post-Newtonian (PN), i.e.~weak 
gravitational field and slow motion, and  post-Minkowskian (PM), i.e.~weak gravitational field, regimes. These predictions build on established perturbative formalisms 
in general relativity \cite{Blanchet:2013haa,Schafer:2018kuf,Futamase:2007zz,Ledvinka:2008tk,Westpfahl:1985}
as well as on the effective field theory approach \cite{Goldberger:2004jt,Foffa:2013gja} yielding a quantum field theoretical diagrammatic 
expansion for the classical effective potential (for introductory reviews see \cite{Goldberger:2007hy,Porto:2016pyg,Levi:2018nxp}).
The present state of the art in the PN expansion is the conservative effective potential at 4PN 
\cite{Damour:2014jta,Damour:2016abl,Bernard:2016wrg,Marchand:2017pir,Foffa:2016rgu,Foffa:2019yfl,Levi:2016ofk}, at 5PN for the static part \cite{Foffa:2019hrb,Blumlein:2019zku} as well as 3PN \cite{Blanchet:2008je} for the gravitational radiation emitted from a quasi-circular inspiral.
In particular the effective-one-body waveform model \cite{Buonanno:1998gg,Buonanno:2000ef,Bohe:2016gbl,Nagar:2018zoe} relies on accurate theoretical predictions for the potential of the binary.

In the past decade important progress in the study of scattering amplitudes in gauge theories
 and
gravity occurred based on innovative perturbative on-shell techniques and discovered mathematical structures that greatly
increased the ability to compute gravitational scattering amplitudes at high perturbative orders. In particular
the efficient double copy construction based on the Bern-Carrasco-Johansson (BCJ) color-kinematical duality \cite{Bern:2008qj, Bern:2010yg, Bern:2010ue}
yields the integrands of gravitational scattering amplitudes from the simpler
quantities in Yang-Mills theories, allowing for high order results in (super)gravity (see e.g.~the recent \cite{Bern:2017puu,Bern:2018jmv}). In view of these innovations it is natural to ask
to what extent these modern scattering amplitude techniques
may be put to work to the classical scattering problem in general relativity  and in turn 
to  the perturbative construction
of effective potentials discussed above. 

The question how the classical two-body gravitational potential  may be extracted 
from the quantum scattering amplitude of (say) massive scalars has a long history 
 \cite{Iwasaki:1971vb,Duff:1973zz,Holstein:2004dn}. Recent works have updated these results by employing the above mentioned modern
unitarity methods for amplitudes \cite{Neill:2013wsa,Bjerrum-Bohr:2013bxa,Luna:2017dtq,Bjerrum-Bohr:2018xdl,Kosower:2018adc}, also including higher curvature terms \cite{Brandhuber:2019qpg,Emond:2019crr},
leading to the 2PM \cite{Cheung:2018wkq,Cristofoli:2019neg} and the first 3PM \cite{Bern:2019nnu} results for the effective gravitational potential very recently.
A 2PM Hamiltonian matched to the classical scattering angle \cite{Westpfahl:1985} was obtained in \cite{Damour:2017zjx,Damour:2016gwp}. 

In parallel there are indications for the existence of a classical double copy of gauge theory
to gravity
beyond the realm of scattering amplitudes. In a series of works \cite{Monteiro:2014cda,Luna:2015paa,Luna:2016hge,Luna:2018dpt,Berman:2018hwd,CarrilloGonzalez:2019gof}
 a selection of gravitational solutions were shown to be double copies of
Yang-Mills ones. Relevant for the problem of classical
gravitational radiation has been the approach of Goldberger and Ridgway \cite{Goldberger:2016iau}
generating perturbative solutions to the equations of motion for a binary pair of spinless massive
particles in dilaton-gravity via the double copy of binary color-charged point particle 
solutions in Yang-Mills theory 
\cite{Goldberger:2017vcg, Goldberger:2017ogt,Goldberger:2017frp,Chester:2017vcz,Bautista:2019tdr}. 
Here color-kinematic replacement rules were employed and
refined in \cite{Shen:2018ebu}, thereby pushing this perturbative double copy  technique to the next-to-next-to-leading order in the coupling constant expansion. In these works
a certain challenge lies in the clean identification of the kinematic numerators and propagators in order to
perform the double copy in analogy to the procedure known from scattering amplitudes, where a factorization into
color factors, kinematic numerators and propagators is manifest. This
approach was lifted from the level of equations of motion to the effective action by
Wormsbecher and two of the present authors \cite{Plefka:2018dpa} recently. Here an adapted version of the double-copy construction was presented, that enabled
a direct computation of the classical effective action for two massive particles in dilaton-gravity
from the corresponding quantity for color charged particles coupled to the Yang-Mills gauge field.
It was shown that this modified prescription to compute the gauge dependent and off-shell effective
potential yields the known result in dilaton-gravity \cite{Damour:1992we} at 2PM level (at the
integrand level) as well as 1PN (explicitly) in a much simpler fashion than the traditional approaches in general relativity. 
This proof of principle thus led us to the hope that it could provide a highly efficient tool to perform
higher order perturbative computations within gravity in a systematic fashion. This motivated the present work to
push this expansion to the next order, i.e.~3PM (at the integrand level) respectively 2PN (exactly).

Unfortunately, we have to report that the effective action generated by the proposed double-copy
procedure of \cite{Plefka:2018dpa} \emph{fails} to agree with the desired dilaton-gravity 
result at this next-to-next-to leading order.
The dilaton-gravity potential at 2PN we sought to reproduce is available in the literature \cite{Mirshekari:2013vb}
and we have also checked its static contributions
from a probe limit comparing to the Janis-Newman-Winicour (JNW)
naked singularity \cite{Janis:1968zz}
as well as through a direct Feynman diagrammatic computation in order to be certain
of the discrepancy. It is important to stress that this breakdown of the
double-copy procedure applies to a gauge variant and off-shell quantity - the effective 
action. This might be the root of the breakdown.
Similar problems should arise at higher orders
for the double copy construction of perturbative spacetimes going beyond the order
considered in \cite{Luna:2016hge}. Note that there as well a comparison to the JNW 
solution occurs, which failed in our scenario.
It is also conceivable that the double copy for scattering amplitudes
involving massive external particles breaks down at the considered order (the leading order
works fine \cite{Luna:2017dtq}).
If this is the case, then problems would also appear for the classical double copy
at the next perturbative order for the emitted radiation at the level of the equations of motion
(i.e.~the order beyond \cite{Shen:2018ebu} which is equivalent to the
the effective action at NLO \cite{Plefka:2018dpa} plus an emitted gluon/graviton accounting for an additional factor of the coupling constant). These interesting questions are left for future
work.

\section{The classical double copy for the effective potential at leading orders in
PM}

The double-copy of pure Yang-Mills theory is the massless sector of bosonic string theory defined by the action
\be
 S_\text{dg} = - \frac{2}{\kappa^2} \int d^4x \sqrt{-g} \left[ R - 2 \partial_\mu \phi \partial^\mu \phi \right]\, ,
\ee
where $\kappa = m_\text{Pl}^{-1} = \sqrt{32\pi G}$ is the gravitational coupling (with Newton's constant $G$ and Planck mass $m_\text{Pl}$), $\phi$ is a real scalar field known as the dilaton. The theory in question also contains an axion
field which will be, however, irrelevant for our considerations. The worldline
action of a point mass $m$ moving along its worldline trajectory $x^{\mu}(\tau)$
reads in the first order formalism (we employ the conventions of \cite{Plefka:2018dpa})
\begin{align}
  S_\text{pm} &= - \int d\tau \left( p_{\mu} \dot x^{\mu} - \lambda(\tau) \left[ e^{-2 \phi} g^{\mu\nu} p_{\mu} p_{\nu} - m^2 \right] \right), \label{eq:YMWL}
\end{align}
where $\lambda$ is a Lagrange multiplier.
The effective potential for two point masses $m$ and $\tilde m$ may be computed in a weak field (or post Minkowskian) expansion by perturbing the metric around a flat Minkowski background 
$g_{\mu\nu}(x) = \eta_{\mu\nu} + \kappa\, h_{\mu\nu}(x)$ and perturbatively integrating out the graviton
field $h_{\mu\nu}$ and the dilaton $\phi$ in the path integral. This yields the effective
action $S_\text{eff,dg}$
\begin{align}
\label{eq3}
  e^{\frac{i}{\hbar} S_\text{eff,dg}} &= e^{\frac{i}{\hbar} S_\text{pm,free}}\mathcal{M}_\text{dg} \nn \\
 & = c\cdot\int \mathcal{D} h \mathcal{D} \phi \, e^{\frac{i}{\hbar} ( S_\text{dg} + S_\text{gf} + S_\text{pm} + \tilde S_\text{pm} )}~,
\end{align}
using a suitable gauge fixing term $S_\text{gf}$.  Here the normalization constant $c$ is chosen
such that $\mathcal{M}_\text{dg}=1$ for $\kappa\to 0$. $S_\text{pm,free}$ is the sum of the 
worldline actions for masses $m$ and $\tilde m$ of \eqn{eq:YMWL} for $g_{\mu\nu}=\eta_{\mu\nu}$ and $\phi=0$.
At leading order in $\kappa$ it is easy to see that $\mathcal{M}_\text{dg}$ 
takes the form 
\begin{fmffile}{Graphs}
\begin{align}  
\raisebox{-0.4cm}{
\begin{fmfgraph*}(40,30)
	\fmfpen{thin}
	\fmfleft{a1,a2}
	\fmfright{b1,b2}
	\fmf{vanilla}{a1,v1,a2}
	\fmf{vanilla}{b1,v2,b2}
	\fmffreeze
	\fmf{dbl_wiggly}{v1,v2}
	\fmfdot{v1,v2}
	\fmfv{label=$\tau_1$}{v1}
	\fmfv{label=$\ttau_1$}{v2}
      \end{fmfgraph*}
     } \quad &= - \frac{i \kappa^2}{\hbar} \int d\hat\tau_{1\tilde 1} \, (p_1 \cdot \tilde p_1)^2 
     D_{1\tilde1} ~ ,
 \end{align}     
with the 4d scalar propagator $D_{ij}:=D(x_{i}-x_{j})$, abbreviating $x_{i}=x(\tau_{i})$
and  with $\Box D(x-y)=-\delta(x-y)$. Moreover we denote
$d\hat \tau := \lambda(\tau) d\tau$, as well as $d\hat\tau_{1\tilde 1}:= d\hat\tau_{1}d\hat\tau_{\tilde 1}$.


The double copy counterpart to \eqn{eq:YMWL} in Yang-Mills theory is a color charged point particle moving along its worldline $x^{\mu}(\tau)$ with color charge $c^{a}(\tau)=\Psi^{\dagger}(\tau)T^{a}\Psi(\tau)$ where $\Psi(\tau)$ is a scalar worldline field.
It couples to the gauge field $A^{a}_{\mu}(x)$ with strength $g$ through the first order action
(for details see \cite{Plefka:2018dpa})
\begin{align}
  S_\text{pc} 
  &= - \int d\tau \Bigl( p_{\mu}  \dot x^{\mu} - i \psi^\dagger \dot \psi \nn \\ & 
            - \lambda \left[ p^2 + 2 g p_{\mu} A^\mu_a c^a + g^2 A_\mu^b c^b A^\mu_a c^a - m^2 \right] \Bigr) .
\end{align}
Note that  the gauge field couples at most quadratically to the worldline, whereas the graviton
has all order couplings in \eqn{eq:YMWL}.
In complete analogy to the gravitational case we define the effective potential for the color charge
$S_\text{eff,YM}$ by integrating out the gluon field in the path integral via
\begin{align}
  e^{\frac{i}{\hbar} S_\text{eff,YM}} &=  e^{\frac{i}{\hbar} S_\text{eff,free}}\mathcal{M}_{\text{YM}} \nn\\ &
  = c'\cdot\int \mathcal{D} A \, e^{\frac{i}{\hbar} (S_\text{YM} + S_\text{gf}+ S_\text{pc} + 
  \tilde S_\text{pc} ) }\, .
\end{align}
The leading order term in $g$ for $\mathcal{M}_\text{YM}$ takes the form
\begin{align}
\raisebox{-0.4cm}{\begin{fmfgraph*}(40,30)
	\fmfpen{thin}
	\fmfleft{a1,a2}
	\fmfright{b1,b2}
	\fmf{vanilla}{a1,v1,a2}
	\fmf{vanilla}{b1,v2,b2}
	\fmffreeze
	\fmf{photon}{v1,v2}
	\fmfdot{v1,v2}
	\fmfv{label=$\tau_1$}{v1}
	\fmfv{label=$\tilde\tau_1$}{v2}
\end{fmfgraph*}}\quad 
&= \frac{4i g^{2}}{\hbar}\int d\hat\tau_{1\tilde1} (c_{1}\cdot \tilde c_{1})\,(p_{1}\cdot \tilde p_{1})\, 
\ D_{1\tilde1}\, .
\label{eq:LOYM}
\end{align} 
Comparing this with \eqn{eq3}
the double-copy structure is obvious: 
Replacing the color factor $(c_{1}\cdot \tilde c_{1})$ by the kinematical numerator
$(p_{1}\cdot \tilde p_{1})$ in \eqn{eq:LOYM} along with the coupling replacement $2g\to i\kappa$ yields the leading order (LO) contribution to the dilaton-gravity effective
potential.

\begin{figure}[t]
\raisebox{-0.4cm}{
\begin{fmfgraph*}(40,30)
	\fmfpen{thin}
	\fmfleft{a1,a2}
	\fmfright{b1,b2}
	\fmf{vanilla}{a1,v1,v3,a2}
	\fmf{vanilla}{b1,v2,v4,b2}
	\fmffreeze
	\fmf{photon}{v1,v2}
	\fmf{photon}{v3,v4}
	\fmfdot{v1,v2,v3,v4}
\end{fmfgraph*}}
\qquad
\raisebox{-0.4cm}{\begin{fmfgraph*}(40,30)
	\fmfpen{thin}
	\fmfleft{a1,a2}
	\fmfright{b1,b2}
	\fmf{vanilla}{a1,v1,a2}
	\fmf{vanilla}{b1,v2,v4,b2}
	\fmffreeze
	\fmf{photon}{v1,v2}
	\fmf{photon}{v1,v4}
	\fmfdot{v1,v2,v4}
	\end{fmfgraph*}}
\qquad
\raisebox{-0.4cm}{
\begin{fmfgraph*}(40,30)
	\fmfpen{thin}
	\fmfleft{a1,a2}
	\fmfright{b1,b2}
	\fmf{vanilla}{a1,v1,a2}
	\fmf{vanilla}{b1,v2,v3,b2}
	\fmffreeze
	\fmf{photon,tension=1.7}{v1,x1}
	\fmf{photon}{x1,v2}
	\fmf{photon}{x1,v3}
	\fmfdot{v1,v2,v3}
	\fmfv{decor.shape=circle,decor.filled=empty,decor.size=0.12w}{x1}
\end{fmfgraph*}}
\caption{Relevant graphs at NLO for the YM effective action.}
\label{fig:YMNLO}
\end{figure}
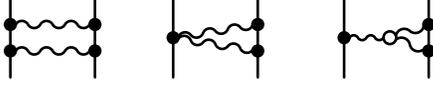

In \cite{Plefka:2018dpa} it was shown that the double-copy prescription
extends to  next-to-leading order (NLO), i.e~$(2g)^{4}\to (i\kappa)^{4}$. Here the relevant  graph topologies on the
Yang-Mills side are collected in FIG.~\ref{fig:YMNLO}. 

The double-copy procedure proposed in \cite{Plefka:2018dpa} amounts to the following steps:
\begin{enumerate}
\item \emph{Transform the YM graphs to a trivalent structure:} For the double gluon couplings to the worldline
this amounts to the replacement
\begin{align}&
\raisebox{-0.5cm}{
		\begin{fmfgraph*}(30,30)
		\fmfpen{thick}
		\fmfleft{s1,s2}
		\fmfright{a,h1,h2,b}
		\fmf{vanilla,width=1}{s1,v1,s2}
		\fmffreeze
		\fmf{photon,width=1}{h1,v1}
		\fmf{photon,width=1}{h2,v1}
		\fmfdot{v1}
		\fmfv{label=$\tau_{1}$,label.dist=2.5}{v1}
		\fmfv{label=$a$ $\mu$,label.dist=2.5}{h1}
		\fmfv{label=$b$ $\nu$,label.dist=2.5}{h2}
		\end{fmfgraph*}} 
	\qquad  	\longrightarrow \quad \delta(\tau_{1}-\tau_{2})\cdot\,\,\,
	\raisebox{-0.5cm}{
		\begin{fmfgraph*}(30,30)
		\fmfpen{thick}
		\fmfleft{s1,s2}
		\fmfright{a,h1,h2,b}
		\fmf{vanilla,width=1}{s1,v1}
		\fmf{dots,width=1}{v1,v2}
		\fmf{vanilla,width=1}{v2,s2}
		\fmffreeze
		\fmf{photon,width=1}{h1,v1}
		\fmf{photon,width=1}{h2,v2}
		\fmfdot{v1,v2}
		\fmfv{label=$\tau_{1}$,label.dist=2.5}{v1}
		\fmfv{label=$\tau_{2}$,label.dist=2.5}{v2}
		\fmfv{label=$a$ $\mu$,label.dist=2.5}{h1}
		\fmfv{label=$b$ $\nu$,label.dist=2.5}{h2}
		\end{fmfgraph*}} \nn
	\\
	&\frac{2i g^{2}}{\hbar}\int d\hat \tau_{1}\, c^{a}(\tau_{1})\, c^{b}(\tau_{1})\, 
	\eta^{\mu\nu} \longrightarrow \label{deltapres}\\ 
	&\frac{2ig^{2}}{\hbar} \int d\hat\tau_{12}  \frac{\delta(\tau_{1}-\tau_{2})}{\lambda(\tau_{2})} 
	\, c^{a}(\tau_{1})\, c^{b}(\tau_{2})\, \eta^{\mu\nu} \, .\nn
\end{align}
In addition, one should seek a color-kinematic duality respecting representation of the bulk graphs dissolving the four-gluon vertex into three-gluon ones.
However, this only arises at the NNLO level and will be discussed in the following. 
\item \emph{Replace color factors by kinematics:} Having established 
the trivalent representation the resulting form of $\mathcal{M}_{\text{YM}}$ takes the general 
form
\begin{equation}
 \qquad \mathcal{M}_{\text{YM}}^{\text{N${}^{n}$LO}} =  (2g)^{2n} 
  \sum_{I\in\Gamma_n} 
  \int\, \prod_{i_{I}}\, d\hat\tau_{i_{I}} \int d^{4l_{I}}x\,
  \frac{C_I\,  N_I}{S_I\, D_I},  \nn
\end{equation}
where $\Gamma_{n}$ represents the set of trivalent graphs at the considered order in perturbation theory, $C_{I}$ the occurring distinct color-factors, $D_{I}$ the associated propagators and
$N_{I}$ the numerators, while $S_{I}$ is the symmetry factor of graph $\Gamma_{i}$. Importantly, we keep the
$\hbar$ dependence in all expressions, i.e. propagators come with a factor of $\hbar/i$
whereas vertices carry a uniform factor of $i/\hbar$.  Once such a representation is established
the double-copy is performed
by simply replacing $C_{I}\to N_{i}$, i.e.
\begin{equation}
\qquad  \mathcal{M}_{\text{dg}}^{\text{N${}^{n}$LO}} =  (i\kappa)^{2n} 
  \sum_{I\in\Gamma_n} 
  \int\, \prod_{i_{I}}\, d\hat\tau_{i_{I}} \int d^{4l_{I}}x\,
  \frac{N_I\,  N_I}{S_I\, D_I} \, , \nn
\end{equation}
which should be the exponentiated effective action of dilaton-gravity.
\item \emph{Establish classical effective action:} To find the classical 
effective potential one  takes the logarithm of $\mathcal{M}_{\text{dg}}$
and sends $\hbar$ to zero, i.e.
\be
\lim_{\hbar\to 0}\, \frac{\hbar}{i}\ln\left [ \sum_{n=0}^{\infty}
 \mathcal{M}_{\text{dg}}^{\text{N${}^{n}$LO}}\right ]=S_{\text{eff,dg}} -  S_\text{pm,free}\, .
 \label{EFAdef}
\ee
In this classical limit ill defined terms arising from squaring $\delta$-functions  in the $N_{I}$ (such as $\frac{\hbar}{i}\delta(0)$) are suppressed. It is expected that they cancel with other quantum
contributions.
\item \emph{Integrate or  perform $PN$ expansion:} Finally, in order to establish the
post-Minkowskian (PM) potential one should perform the bulk and $\tau_{i}$ integrals. For the
post-Newtonian  (PN) approximation one first takes this limit and thereafter integrates.
\end{enumerate}

The resulting effective action from \eqn{EFAdef} was shown to agree with the result
in the literature on scalar-tensor theories  \cite{Damour:1992we}  up to and including 1PN order. Moreover, it was shown in \cite{Plefka:2018dpa} that the double-copy result also agrees at the 2PM order
at the level of integrands.

\section{Yang-Mills computation at NNLO}


%
%
%
We now turn to the discussion of the NNLO computation via the double copy.
To begin with consider the group of the diagrams that, with the introduction of the appropriate delta functions in the sense of \eqn{deltapres}, share the color factor of the three 
ladder diagram, i.e.~$(c\cdot\tilde c)^{3}$. 
\begin{widetext}
\begin{align}
\raisebox{-0.4cm}
{
\begin{fmfgraph}(40,30)
	\fmfpen{thin}
	\fmfleft{a1,a2}
	\fmfright{b1,b2}
	\fmf{vanilla}{a1,v1,v2,v3,a2}
	\fmf{vanilla}{b1,u1,u2,u3,b2}
	\fmffreeze
	\fmf{photon}{v1,u1}
	\fmf{photon}{v2,u2}
	\fmf{photon}{v3,u3}
	\fmfdot{v1,v2,v3,u1,u2,u3}
	\fmfv{label=$\tau_1$}{v1}
	\fmfv{label=$\tau_2$}{v2}
	\fmfv{label=$\tau_3$}{v3}
	\fmfv{label=$\tilde\tau_1$}{u1}
	\fmfv{label=$\tilde\tau_2$}{u2}
	\fmfv{label=$\tilde\tau_3$}{u3}
\end{fmfgraph}
 } &= {1\over 6} \left({i\over \hbar} \right)^3 (2g)^6 \int d\hat \tau_{123\tilde 1\tilde 2\tilde 3} (p_1\cdot \tilde p_1) (p_2\cdot \tilde p_2) (p_3 \cdot \tilde p_3) (c_1\cdot \tilde c_1)(c_2\cdot \tilde c_2)(c_3\cdot \tilde c_3) D_{1\tilde 1} D_{2\tilde 2} D_{3 \tilde 3}
 \label{eq10}\\
\raisebox{-0.4cm}
{
\begin{fmfgraph}(40,30)
	\fmfpen{thin}
	\fmfleft{a1,a2}
	\fmfright{b1,b2}
	\fmf{vanilla}{a1,v1,v2,a2}
	\fmf{vanilla}{b1,u1,u2,u3,b2}
	\fmffreeze
	\fmf{photon}{v1,u1}
	\fmf{photon}{v1,u2}
	\fmf{photon}{v2,u3}
	\fmfdot{v1,v2,u1,u2,u3}
	\fmfv{label=$1$}{v1}
	\fmfv{label=$3$}{v2}
	\fmfv{label=$\tilde 1$}{u1}
	\fmfv{label=$\tilde 2$}{u2}
	\fmfv{label=$\tilde 3$}{u3}
\end{fmfgraph}
} 
& =  {1\over 6} \left({i\over \hbar} \right)^3 (2g)^6 \int d\hat \tau_{123\tilde 1\tilde 2\tilde 3} {1\over 2} {\hbar \over i }\Big[ { \delta (\tau_1-\tau_2) \over \lambda_2 } (\tilde p_1\cdot \tilde p_2) (p_3\cdot \tilde p_3) + { \delta (\tau_2-\tau_3) \over \lambda_3 } (\tilde p_2\cdot \tilde p_3) (p_1\cdot \tilde p_1) \nonumber \\
& \qquad\qquad\qquad \qquad\qquad\qquad \qquad +  { \delta (\tau_3-\tau_1) \over \lambda_1 } (\tilde p_1\cdot \tilde p_3) (p_2\cdot \tilde p_2) \Big] (c_1\cdot \tilde c_1)(c_2\cdot \tilde c_2)(c_3\cdot \tilde c_3) D_{1\tilde 1} D_{2\tilde 2} D_{3 \tilde 3} 
\label{eq11} \\
\raisebox{-0.4cm}
{
\begin{fmfgraph}(40,30)
	\fmfpen{thin}
	\fmfleft{a1,a2}
	\fmfright{b1,b2}
	\fmf{vanilla}{a1,v1,v2,a2}
	\fmf{vanilla}{b1,u1,u2,b2}
	\fmffreeze
	\fmf{photon}{v1,u1}
	\fmf{photon}{v2,u2}
	\fmf{photon}{v1,u2}
	\fmfdot{v1,v2,u1,u2}
	\fmfv{label=$1$}{v1}
	\fmfv{label=$3$}{v2}
	\fmfv{label=$\tilde 1$}{u1}
	\fmfv{label=$\tilde 3$}{u2}
\end{fmfgraph}
}  &= {1\over 6} \left({i\over \hbar} \right)^3 (2g)^6  \int d\hat \tau_{123\tilde 1\tilde 2\tilde 3} {1 \over 6} \left( {\hbar \over i} \right)^2 \Big[ {\delta(\tau_1-\tau_2) \over \lambda_2} p_3^\mu + \text{cyclic}(1, 2, 3) \Big] \Big[ {\delta(\tilde \tau_1-\tilde \tau_2)\over \tilde \lambda_2} \tilde p_3^\nu + \text{cyclic}(\tilde 1, \tilde 2, \tilde 3) \Big] \eta_{\mu\nu} \nonumber \\ 
& \qquad \qquad \qquad \qquad \qquad \qquad \qquad \times (c_1\cdot \tilde c_1)(c_2\cdot \tilde c_2)(c_3\cdot \tilde c_3) D_{1\tilde 1} D_{2 \tilde 2} D_{3 \tilde 3} \label{eq12}
\end{align}
Note that there is also a mirrored graph to \eqn{eq11} obtained by swapping tilded and non-tilded quantities. Also, we write the integrands always in a completely 
symmetrized fashion with respect to permutations of the $\tau_{i}$ and $\tilde\tau_{i}$
proper-time variables. This prescription has an impact on the double copy and guarantees
exponentiation as we shall see.
The double copy of these diagrams will have a nontrivial contribution at 2PN.

The second group of diagrams shares one bulk vertex and has the common color factor $(c\cdot [\tilde c, \tilde c]) (c\cdot
\tilde c)$ reading
\begin{align}
\raisebox{-0.4cm}
{
\begin{fmfgraph}(40,30)
	\fmfpen{thin}
	\fmfleft{a1,a2}
	\fmfright{b1,b2}
	\fmf{vanilla}{a1,v1,v2,a2}
	\fmf{vanilla}{b1,u1,u2,u3,b2}
	\fmffreeze
	\fmf{photon,tension=2.7}{v1,x}
	\fmf{photon,tension=1.5}{x,u1}
	\fmf{photon,tension=1.5}{x,u2}
	\fmf{photon}{v2,u3}
	\fmfdot{v1,v2,u1,u2,u3}
	\fmfv{decor.shape=circle,decor.filled=empty,decor.size=0.1w}{x}
\end{fmfgraph}
} & = {1\over 2} \left( {i\over \hbar} \right)^2 (2g)^6 \int d\hat\tau_{13\tilde 1\tilde 2\tilde 3}\; \left[ -{1\over 2}V^{\mu\nu\rho}_{1\tilde 1} p_{1\mu} \tilde p_{ 1\nu} \tilde p_{2\rho} \right] (p_3 \cdot \tilde p_{3}) f^{abc} c_{1a} \tilde c_{1b} \tilde c_{2c} (c_3\cdot \tilde c_3) G_{1\tilde 1\tilde 2} D_{3\tilde 3} \\
\raisebox{-0.4cm}
{
\begin{fmfgraph}(40,30)
	\fmfpen{thin}
	\fmfleft{a1,a2}
	\fmfright{b1,b2}
	\fmf{vanilla}{a1,v1,v2,a2}
	\fmf{vanilla}{b1,u1,u2,b2}
	\fmffreeze
	\fmf{photon,tension=2.7}{v1,x}
	\fmf{photon,tension=1.5}{x,u1}
	\fmf{photon,tension=1.5}{x,u2}
	\fmf{photon}{v2,u2}
	\fmfdot{v1,v2,u1,u2}
	\fmfv{decor.shape=circle,decor.filled=empty,decor.size=0.1w}{x}
\end{fmfgraph}
} &=  {1\over 2} \left({i\over \hbar}\right)^2 (2g)^6 \int d\hat\tau_{13\tilde 1\tilde 2\tilde 3}\; {\hbar\over i} \left[  -{1\over 4} {\delta (\tilde\tau_2 -\tilde\tau_3) \over \tilde\lambda_3}  V^{\mu\nu\rho}_{1\tilde 1} \tilde p_{1\nu} + ( \tilde 1 \leftrightarrow \tilde 2) \right]p_{1\mu} p_{3\rho} f^{abc} c_{1a} \tilde c_{1b} \tilde c_{2c} (c_3\cdot \tilde c_3) G_{1\tilde 1\tilde 2}D_{3\tilde 3} \\
\raisebox{-0.4cm}
{
\begin{fmfgraph}(40,30)
	\fmfpen{thin}
	\fmfleft{a1,a2}
	\fmfright{b1,b2}
	\fmf{vanilla}{a1,v1,a2}
	\fmf{vanilla}{b1,u1,u2,u3,b2}
	\fmffreeze
	\fmf{photon,tension=2.7}{v1,x}
	\fmf{photon,tension=1.5}{x,u1}
	\fmf{photon,tension=1.5}{x,u2}
	\fmf{photon}{v1,u3}
	\fmfdot{v1,u1,u2,u3}
	\fmfv{decor.shape=circle,decor.filled=empty,decor.size=0.1w}{x}
\end{fmfgraph}
} &= {1\over 2} \left({i\over \hbar}\right)^2 (2g)^6 \int d\hat\tau_{13\tilde 1\tilde 2\tilde 3}\; {\hbar \over i} {\delta(\tau_1-\tau_3) \over \lambda_3} f^{abc} c_{1a} \tilde c_{1b} \tilde c_{2c} (c_3\cdot \tilde c_3) \left[ -{1\over 4}V^{\mu\nu\rho}_{1\tilde 1}\tilde p_{ 3\mu} \tilde p_{ 1\nu} \tilde p_{ 2 \rho} \right] G_{1\tilde 1\tilde 2} D_{3\tilde 3}
\end{align}
where
$
V_{12}^{\mu\nu\rho} = \eta^{\mu\nu}\left( \partial_1 - \partial_2 \right)^{\rho} + \eta^{\nu\rho}\left( \partial_1 + 2 \partial_2 \right)^{\mu}  + \eta^{\rho\mu}\left(-2 \partial_1 - \partial_2 \right)^{\nu} $ and $G_{1\tilde 1\tilde 2} =\int d^4 x D_{1x} D_{\tilde 1 x} D_{\tilde 2 x}$.
It turns out that the double-copy of these graphs is relevant to the 3PM, but does not
contribute at the 2PN level. Here we only display their 3PM integrals and 
suppressed the mirror diagrams obtained by swapping the tilded and non-tilded indices.

The third group of diagrams containing bulk vertices at order $g^{2}$ being symmetric
with respect to both worldlines is characterized by
the color structures $[c, c]^{a}\, [\tilde c ,\tilde c]^{a}$
and $[c, \tilde c]^{a}\, [\tilde c , c]^{a}$. One now has
\begin{align}
\raisebox{-0.4cm}
{
\begin{fmfgraph}(40,30)
	\fmfpen{thin}
	\fmfleft{a1,a2}
	\fmfright{b1,b2}
	\fmf{vanilla}{a1,v1,v2,a2}
	\fmf{vanilla}{b1,u1,u2,b2}
	\fmffreeze
	\fmf{photon,tension=1.5}{v1,x}
	\fmf{photon,tension=1.5}{v2,x}
	\fmf{photon,tension=2.5}{x,y}
	\fmf{photon,tension=1.5}{y,u1}
	\fmf{photon,tension=1.5}{y,u2}
	\fmfdot{v1,v2,u1,u2}
	\fmfv{decor.shape=circle,decor.filled=empty,decor.size=0.1w}{x,y}
\end{fmfgraph}
} & ={1\over 4} {i\over \hbar} (2g)^6 \int d\hat\tau_{12\tilde 1\tilde 2}\; f^{abe} f^{cde} c_{1a} c_{2b} \tilde c_{1c} \tilde c_{2d} \left[ {1\over 4}  V^{\mu\nu\lambda}_{12}  V^{\rho\sigma\delta}_{\tilde 1\tilde 2} \eta_{\lambda\delta} p_{1\mu} p_{2\nu} \tilde  p_{1\rho} \tilde  p_{2\sigma} \right] G_{12 ;\tilde1\tilde 2} \label{eq16} \\
\raisebox{-0.4cm}
{
\begin{fmfgraph}(40,30)
	\fmfpen{thin}
	\fmfleft{a1,a2}
	\fmfright{b1,b2}
	\fmf{vanilla}{a1,v1,v2,a2}
	\fmf{vanilla}{b1,u1,u2,b2}
	\fmffreeze
	\fmf{photon,tension=1.5}{v1,x}
	\fmf{photon,tension=1.5}{x,u1}
	\fmf{photon,tension=0.3}{x,y}
	\fmf{photon,tension=1.5}{v2,y}
	\fmf{photon,tension=1.5}{y,u2}
	\fmfdot{v1,v2,u1,u2}
	\fmfv{decor.shape=circle,decor.filled=empty,decor.size=0.1w}{x,y}
\end{fmfgraph}
} &= {\xi\over 2} {i\over \hbar} (2g)^6 \int d\hat\tau_{12\tilde 1\tilde 2}\; f^{ace} f^{bde} c_{1a} c_{2b} \tilde c_{1c} \tilde c_{2d} \left[ {1\over 4}  V^{\mu\rho\lambda}_{1\tilde 1} V^{\nu\sigma\delta}_{2\tilde 2} \eta_{\lambda\delta} p_{1\mu} p_{2\nu} \tilde  p_{1\rho} \tilde  p_{2\sigma} \right] G_{1\tilde 1 ;2\tilde 2} \label{tcd}\\
\raisebox{-0.4cm}
{
\begin{fmfgraph}(40,30)
	\fmfpen{thin}
	\fmfleft{a1,a2}
	\fmfright{b1,b2}
	\fmf{vanilla}{a1,v1,v2,a2}
	\fmf{vanilla}{b1,u1,u2,b2}
	\fmffreeze
	\fmf{photon,tension=1.2}{v1,x}
	\fmf{photon,tension=1.2}{v2,y}
	\fmf{photon,tension=0}{x,y}
	\fmf{phantom}{x,u1}
	\fmf{phantom}{y,u2}
	\fmf{photon,tension=0.2}{x,u2}
	\fmf{photon,tension=0.2}{y,u1}
	\fmfdot{v1,v2,u1,u2}
	\fmfv{decor.shape=circle,decor.filled=empty,decor.size=0.1w}{x,y}
\end{fmfgraph}
} &= {1-\xi\over 2} {i\over \hbar} (2g)^6 \int d\hat\tau_{12\tilde 1\tilde 2}\; f^{ade} f^{bce} c_{1a} c_{2b} \tilde c_{1c} \tilde c_{2d} \left[ {1\over 4}  V^{\mu\sigma\lambda}_{1\tilde 2}  V^{\nu\rho\delta}_{2\tilde 1} \eta_{\lambda\delta} p_{1\mu} p_{2\nu} \tilde  p_{1\rho} \tilde  p_{2\sigma} \right] G_{1\tilde 2 ;2\tilde 1} \label{ucd} \\
\raisebox{-0.4cm}
{
\begin{fmfgraph}(40,30)
	\fmfpen{thin}
	\fmfleft{a1,a2}
	\fmfright{b1,b2}
	\fmf{vanilla}{a1,v1,v2,a2}
	\fmf{vanilla}{b1,u1,u2,b2}
	\fmffreeze
	\fmf{photon}{v1,x}
	\fmf{photon}{x,u1}
	\fmf{photon}{v2,x}
	\fmf{photon}{x,u2}
	\fmfdot{v1,v2,u1,u2}
	\fmfv{decor.shape=circle,decor.filled=empty,decor.size=0.1w}{x}
\end{fmfgraph}
} &= {1\over 4} {i\over \hbar} (2g)^6 \int d\hat\tau_{12\tilde 1\tilde 2} {1\over 4}\Big[ f^{abe} f^{cde} c_{1a} c_{2b} \tilde c_{1c} \tilde c_{2d} \left[ (p_1\cdot \tilde p_{1}) (p_2\cdot \tilde p_{2}) - (p_1\cdot \tilde p_{2}) (p_2\cdot \tilde p_{1})  \right] (\partial_1+\partial_2)^2 G_{12;\tilde 1\tilde 2}  \nonumber\\
& \qquad\qquad \quad \;+ 2\rho f^{ace} f^{bde} c_{1a} c_{2b} \tilde c_{1c} \tilde c_{2d} \left[ (p_1 \cdot p_2) ( \tilde p_1\cdot \tilde p_2) - (p_1\cdot \tilde p_2) ( p_2 \cdot \tilde p_1) \right] (\partial_1 + \tilde\partial_1)^2 G_{1\tilde 1;2\tilde 2} \nonumber\\
&  \qquad\qquad \quad \;+2(1-\rho) f^{ade} f^{bce} c_{1a} c_{2b} \tilde c_{1c} \tilde c_{2d} \left[ (p_1\cdot p_2)  (\tilde p_1\cdot \tilde p_2) - (p_1\cdot \tilde p_1) ( p_2 \cdot \tilde p_2) \right] (\partial_1 + \tilde\partial_2)^2 G_{1\tilde 2;2\tilde 1}\Big],  \label{eq19}
\end{align}
where we have introduced the two-loop function $G_{12;\tilde 1\tilde 2} =\int d^4 x\, d^{4}y  D_{1x} D_{2 x}  D_{xy}D_{\tilde 1 y}D_{\tilde 2 y}$. Note that the diagrams 
\eqn{tcd} and \eqn{ucd} are numerically identical under a relabeling of the
worldline variables $\tilde\tau_{i}$. However, they will lead to different double copies. This ambiguity is captured by introducing the parameter $\xi$. The same applies to the last two lines
of \eqn{eq19}, whose ambiguity we parametrize with $\rho$. 
Of course, the right choice should be 
dictated by requiring the color-kinematic algebra, i.e.~kinematical Jacobi identity,
to hold for the numerators -- thereby possibly
adding vanishing terms to these expressions, reflecting generalized gauge transformations
\cite{Bern:2010yg}. 
We will postpone this analysis to the post-Newtonian
limit in the next section.

The final group of bulk diagrams is non-symmetric in the two worldlines and
carries the color structures $[c,c]^a [c,\tilde c]^a$ and
mirrors $[c,\tilde c]^a [\tilde c,\tilde c]^a$. They also contribute to 2PN and read
\begin{align}
\raisebox{-0.4cm}
{
\begin{fmfgraph}(40,30)
	\fmfpen{thin}
	\fmfleft{a1,a2}
	\fmfright{b1,b2}
	\fmf{vanilla}{a1,v1,v2,v3,a2}
	\fmf{vanilla}{b1,u1,b2}
	\fmffreeze
	\fmf{photon,tension=0.1}{v1,x}
	\fmf{photon,tension=0.1}{v2,x}
	\fmf{photon,tension=0.2}{x,y}
	\fmf{photon}{v3,y}
	\fmf{photon,tension=1.5}{y,u1}
	\fmfdot{v1,v2,v3,u1}
	\fmfv{decor.shape=circle,decor.filled=empty,decor.size=0.1w}{x,y}
\end{fmfgraph}
} = & {1\over 6}{i\over \hbar} (2g)^6 \int d\hat\tau_{123\tilde 1} {1\over 4} \Big[ \alpha_1  f^{abe} f^{cde} c_{1a} c_{2b} c_{3c} \tilde c_{1d} V^{\mu\nu\lambda}_{12}  V^{\rho\sigma\delta}_{3\tilde 1} \eta_{\lambda\delta} p_{1\mu} p_{2\nu} p_{3\rho} \tilde p_{1\sigma} G_{12; 3\tilde 1} \nonumber \\
&\qquad\qquad \qquad \qquad + \alpha_2  f^{ace} f^{bde} c_{1a} c_{2b} c_{3c} \tilde c_{1d} V^{\mu\rho\lambda}_{13}  V^{\nu\sigma\delta}_{2\tilde 1} \eta_{\lambda\delta} p_{1\mu} p_{2\nu} p_{3\rho} \tilde p_{1\sigma} G_{13; 2\tilde 1} \nonumber \\
&\qquad\qquad \qquad \qquad + (3-\alpha_1-\alpha_2)  f^{ade} f^{bce} c_{1a} c_{2b} c_{3c} \tilde c_{1d} V^{\mu\sigma\lambda}_{1\tilde 1}  V^{\nu\rho\delta}_{23} \eta_{\lambda\delta} p_{1\mu} p_{2\nu} p_{3\rho} \tilde p_{1\sigma} G_{1\tilde 1; 23} \Big] \label{eq20} \\
\raisebox{-0.4cm}
{
\begin{fmfgraph}(40,30)
	\fmfpen{thin}
	\fmfleft{a1,a2}
	\fmfright{b1,b2}
	\fmf{vanilla}{a1,v1,v2,v3,a2}
	\fmf{vanilla}{b1,u1,b2}
	\fmffreeze
	\fmf{photon,tension=0.1}{v1,x}
	\fmf{photon,tension=0.1}{v2,x}
	\fmf{photon,tension=0.1}{v3,x}
	\fmf{photon,tension=0.3}{x,u1}
	\fmfdot{v1,v2,v3,u1}
	\fmfv{decor.shape=circle,decor.filled=empty,decor.size=0.1w}{x}
\end{fmfgraph}
} = & {1 \over 6} {i\over \hbar} (2g)^6 \int d \hat\tau_{123\tilde 1} {1\over 4}\Big[\beta_1 f^{abe} f^{cde} c_{1a} c_{2b} c_{3c} \tilde c_{1d} \left[ (p_1\cdot p_3) (p_2\cdot \tilde p_{1}) - (p_1\cdot \tilde p_{1}) (p_2\cdot p_{3})  \right] (\partial_1+\partial_2)^2 G_{12;3\tilde 1}  \nonumber\\
& \qquad\  + \beta_2 f^{ace} f^{bde} c_{1a} c_{2b} c_{3c} \tilde c_{1d} \left[ (p_1 \cdot p_2) ( p_3\cdot \tilde p_1) - (p_1\cdot \tilde p_1) ( p_2 \cdot p_3) \right] (\partial_1 + \partial_3)^2 G_{13;2\tilde 1} \nonumber\\
& \qquad\ + ( 3 - \beta_1 - \beta_2 ) f^{ade} f^{bce} c_{1a} c_{2b} c_{3c} \tilde c_{1d} \left[ (p_1\cdot p_2)  (p_3\cdot \tilde p_1) - (p_1\cdot p_3) ( p_2 \cdot \tilde p_1) \right] (\partial_1 + \tilde\partial_1)^2 G_{1\tilde 1; 32}\Big]. \label{eq21}
\end{align}
\end{widetext}
Again, we use $\alpha_1$, $\alpha_2$, $\beta_1$, $\beta_2$ to parametrize the ambiguity of relabeling the worldline variables $\tau_i$. Naturally, there is also the set of
mirrored graphs swapping tilded and non-tilded quantities.

\section{Double copy prescription at NNLO}

\begin{figure}[b]
\begin{center}
\begin{fmfgraph}(40,32)
	\fmfpen{thin}
	\fmfleft{a1,a2}
	\fmfright{b1,b2}
	\fmf{vanilla}{a1,v1,a2}
	\fmf{vanilla}{b1,u1,u2,u3,b2}
	\fmffreeze
	\fmf{vanilla}{v1,u1}
	\fmf{vanilla}{v1,u2}
	\fmf{vanilla}{v1,u3}
	\fmfdot{v1,u1,u2,u3}
\end{fmfgraph}
\quad
\begin{fmfgraph}(40,32)
	\fmfpen{thin}
	\fmfleft{a1,a2}
	\fmfright{b1,b2}
	\fmf{vanilla}{a1,v1,v2,a2}
	\fmf{vanilla}{b1,u1,u2,b2}
	\fmffreeze
	\fmf{vanilla}{v1,u1}
	\fmf{vanilla}{v1,u2}
	\fmf{vanilla}{v2,u2}
	\fmfdot{v1,v2,u1,u2}
\end{fmfgraph}
\quad
\begin{fmfgraph}(40,32)
	\fmfpen{thin}
	\fmfleft{a1,a2}
	\fmfright{b1,b2}
	\fmf{vanilla}{a1,v1,a2}
	\fmf{vanilla}{b1,u1,u2,u3,b2}
	\fmffreeze
	\fmf{vanilla}{v1,x}
	\fmf{vanilla}{x,u1}
	\fmf{vanilla}{x,u2}
	\fmf{vanilla}{v1,u3}
	\fmfdot{v1,u1,u2,u3}
	\fmfv{decor.shape=circle,decor.filled=empty,decor.size=0.1w}{x}
\end{fmfgraph}
\quad
\begin{fmfgraph}(40,32)
	\fmfpen{thin}
	\fmfleft{a1,a2}
	\fmfright{b1,b2}
	\fmf{vanilla}{a1,v1,v2,a2}
	\fmf{vanilla}{b1,u1,u2,b2}
	\fmffreeze
	\fmf{vanilla}{v1,x}
	\fmf{vanilla}{v2,u2}
	\fmf{vanilla}{x,u1}
	\fmf{vanilla}{x,u2}
	\fmfdot{v1,v2,u1,u2}
	\fmfv{decor.shape=circle,decor.filled=empty,decor.size=0.1w}{x}
\end{fmfgraph}
\end{center}
\begin{center}
\begin{fmfgraph}(40,32)
	\fmfpen{thin}
	\fmfleft{a1,a2}
	\fmfright{b1,b2}
	\fmf{vanilla}{a1,v1,v2,a2}
	\fmf{vanilla}{b1,u1,u2,b2}
	\fmffreeze
	\fmf{vanilla}{v1,x}
	\fmf{vanilla}{v2,x}
	\fmf{vanilla}{x,y}
	\fmf{vanilla}{y,u1}
	\fmf{vanilla}{y,u2}
	\fmfdot{v1,v2,u1,u2}
	\fmfv{decor.shape=circle,decor.filled=empty,decor.size=0.1w}{x,y}
\end{fmfgraph}
\quad
\begin{fmfgraph}(40,32)
	\fmfpen{thin}
	\fmfleft{a1,a2}
	\fmfright{b1,b2}
	\fmf{vanilla}{a1,v1,v2,a2}
	\fmf{vanilla}{b1,u1,u2,b2}
	\fmffreeze
	\fmf{vanilla}{v1,x}
	\fmf{vanilla}{x,u1}
	\fmf{vanilla,tension=0.1}{x,y}
	\fmf{vanilla}{v2,y}
	\fmf{vanilla}{y,u2}
	\fmfdot{v1,v2,u1,u2}
	\fmfv{decor.shape=circle,decor.filled=empty,decor.size=0.1w}{x,y}
\end{fmfgraph}
\quad
\begin{fmfgraph}(40,32)
	\fmfpen{thin}
	\fmfleft{a1,a2}
	\fmfright{b1,b2}
	\fmf{vanilla}{a1,v1,v2,a2}
	\fmf{vanilla}{b1,u1,u2,b2}
	\fmffreeze
	\fmf{vanilla}{v1,x}
	\fmf{vanilla}{v2,x}
	\fmf{vanilla}{x,u1}
	\fmf{vanilla}{x,u2}
	\fmfdot{v1,v2,u1,u2}
	\fmfv{decor.shape=circle,decor.filled=empty,decor.size=0.1w}{x}
\end{fmfgraph}
\quad
\begin{fmfgraph}(40,32)
	\fmfpen{thin}
	\fmfleft{a1,a2}
	\fmfright{b1,b2}
	\fmf{vanilla}{a1,v1,a2}
	\fmf{vanilla}{b1,u1,u2,u3,b2}
	\fmffreeze
	\fmf{vanilla,tension=1.5}{v1,x}
	\fmf{vanilla,tension=0.2}{x,y}
	\fmf{vanilla,tension=0.1}{y,u1}
	\fmf{vanilla,tension=0.1}{y,u2}
	\fmf{vanilla}{x,u3}
	\fmfdot{v1,u1,u2,u3}
	\fmfv{decor.shape=circle,decor.filled=empty,decor.size=0.1w}{x,y}
\end{fmfgraph}
\quad
\begin{fmfgraph}(40,32)
	\fmfpen{thin}
	\fmfleft{a1,a2}
	\fmfright{b1,b2}
	\fmf{vanilla}{a1,v1,a2}
	\fmf{vanilla}{b1,u1,u2,u3,b2}
	\fmffreeze
	\fmf{vanilla,tension=0.3}{v1,x}
	\fmf{vanilla,tension=0.1}{x,u1}
	\fmf{vanilla,tension=0.1}{x,u2}
	\fmf{vanilla,tension=0.1}{x,u3}
	\fmfdot{v1,u1,u2,u3}
	\fmfv{decor.shape=circle,decor.filled=empty,decor.size=0.1w}{x}
\end{fmfgraph}
\end{center}

\caption{Topologies generated by the double copy at NNLO.}
\label{fig:DCgraphs}
\end{figure}
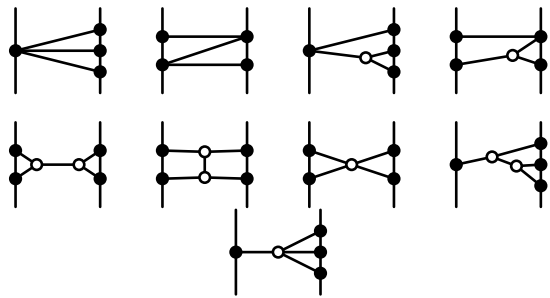

In order to clearly spell out the double copy prescription for NNLO that we applied we do this in greater detail
for the first group of diagrams
with color structure $(c\cdot \tilde c)^{3}$ from \eqn{eq10}--\eqn{eq12}. 
The sum of these graphs takes the form 
\begin{align}
\mathcal{M}_{\text{YM}}^{(c\cdot\tilde c)^{3}}&= \label{eq22} \\
 \frac 16\left(\frac i \hbar \right)^3 & (2g)^6  \int d\hat \tau_{123\tilde 1\tilde 2\tilde 3}\; 
 N_{(c\cdot\tilde c)^{3}}  \, C_{(c\cdot\tilde c)^{3}}\, D_{1\tilde 1} D_{2 \tilde 2} D_{3 \tilde 3}
 \nonumber
 \end{align}
with the color factor
$
 C_{(c\cdot\tilde c)^{3}}=(c_1\cdot \tilde c_1)(c_2\cdot \tilde c_2)(c_3\cdot \tilde c_3)
$
and the kinematic numerator
\begin{gather}
 N_{(c\cdot\tilde c)^{3}}=
 (p_1\cdot \tilde p_1) (p_2\cdot \tilde p_2) (p_3 \cdot \tilde p_3) \\
 + 
 {1\over 2} {\hbar \over i }\Big[ { \delta (\tau_1-\tau_2) \over \lambda_2 } (\tilde p_1\cdot \tilde p_2) (p_3\cdot \tilde p_3) +  \text{cyclic} (1,2,3) \nonumber \\
 \qquad + { \delta (\tilde\tau_1-\tilde\tau_2) \over \tilde\lambda_2 } (p_1\cdot p_2) ( p_3\cdot \tilde p_3) + \text{cyclic} (\tilde 1, \tilde 2, \tilde 3)\Big] \nonumber \\
 +{3\over 2} \left( {\hbar \over i} \right)^2 {\delta(\tau_1-\tau_2) \over \lambda_2} {\delta(\tilde \tau_3-\tilde \tau_2)\over \tilde \lambda_2} (p_2\cdot \tilde p_1)
\end{gather}
The double copy is then performed by replacing $ C_{(c\cdot\tilde c)^{3}}\rightarrow N_{(c\cdot\tilde c)^{3}}$ and $2g\to i\kappa$ in \eqn{eq22} thereby constructing the putative
dilaton-gravity contribution
 $\mathcal{M}_{\text{dg}}^{(c\cdot\tilde c)^{3}} $ from Yang-Mills theory. 
The obtained expressions indeed exponentiate:
\begin{widetext}
\begin{equation}
\begin{split}
\mathcal{M}_{\text{dg}}^{(c\cdot\tilde c)^{3}} &= \exp\Bigl [
\frac{i}{\hbar} \frac{(i\kappa)^{2}}{2}\int d\hat \tau_{1\tilde 2}  (p_{1}\cdot \tilde p_{2})^{2} D_{1\tilde 2}
+ \frac{i}{\hbar} \frac{(i\kappa)^{4}}{2} \int d\hat \tau_{1\tilde 2\tilde 3} (p_{1}\cdot \tilde p_{2})
(p_{1}\cdot \tilde p_{3})(\tilde p_{2}\cdot \tilde p_{3}) D_{1\tilde 2}D_{1\tilde 3} 
 \\
		&+ \frac{i}{\hbar} \frac{(i\kappa)^{6}}{2}  \int d\hat\tau_{12\tilde 1\tilde 2} {1\over 2}\left[ (p_1\cdot \tilde{p}_{ 1}) (p_2\cdot {\tilde p_2}) (p_1\cdot {\tilde p_ 2} ) (p_2 \cdot {\tilde p_ 1}) +  (p_1\cdot {\tilde p_1}) (p_2\cdot {\tilde p_2}) (p_1\cdot p_2) ({\tilde p_1} \cdot {\tilde p_2})  \right]D_{1\tilde 1} D_{2\tilde 2} D_{2\tilde 1} \\ &
		+\frac{i}{\hbar} (i\kappa)^{6}  \int d\hat\tau_{1\tilde 1\tilde 2\tilde 3} {1\over 4} (p_1\cdot {\tilde p_1}) (p_1\cdot {\tilde p_3}) ({\tilde p_1}\cdot {\tilde p_3}) ({\tilde p_2}\cdot {\tilde p_3}) D_{1\tilde 1} D_{1\tilde 2} D_{1\tilde 3} + \text{(mirrored)}
+ \cO(\hbar^{0}) \, \Bigr ] \, \Bigr |_{\kappa^{6}}\,.
\label{eq26}
\end{split}
\end{equation}
\end{widetext}
Note that the first and third term in the above is mirror symmetric, hence the factor of $\ft 12$.
Importantly also the suppressed quantum terms at $\cO(\hbar ^{0})$ exponentiate, which is only true if one symmetrizes all the $\tau_{i}$ integrands in the YM-representation. They contain ill defined expressions
proportional to $\delta(0)$. We hence consistently recover the exponentiated 
LO and NLO order results of \cite{Plefka:2018dpa} at this order in $\kappa^{6}$. 
The contributions at NNLO to the effective action from this sector are thus given by the last two lines in \eqn{eq26}. 
Performing 
the double copy of the other groups of graphs, i.e.~the
symmetric bulk graphs stemming from eqs.~\eqn{eq16}--\eqn{eq19} as well as the non-symmetric
bulk graphs arising from eqs.~\eqn{eq20}--\eqn{eq21} proceeds along the same lines. 
Under the double copy one then produces the NNLO topologies displayed in FIG.~\ref{fig:DCgraphs}. They indeed match the topologies present in gravity at the 2PN
order, see e.g.~\cite{Gilmore:2008gq}.
The total effective action at 3PM order produced via the double copy
would follow by performing the worldline and bulk integrals. We shall, however, proceed to
work out the 2PN expansion whose integrals are more straightforward to obtain.

Finally, we remark that the double copy prescription that we applied here, notably the
rewriting of the worldline-bulk vertices in trivalent ones using $\delta$-functions, also
follows from a systematic treatment employing
the bi-fundamental scalar field theory as the seed to define the kinematic numerators and denominators.
This route was advocated in \cite{Shen:2018ebu} for the analogous radiation
problem at the level of the equations of motion. This fact essentially hinges on the simple observation
that the worldline only interacts linearly with the bulk fields in bi-fundamental scalar field theory also in the first-order worldline formalism.

\section{PN expansion and kinematical Jacobi identity}

%
%

Details of performing the post-Newtonian (PN) expansion were discussed
in \cite{Plefka:2018dpa}.  The expansion 
combines a weak-field \emph{and} slow-motion approximation for bound
binaries. The virial theorem in this setup states
\begin{equation}
  \frac{\vct{v}_r^2}{c^2} \sim \frac{\kappa^2 (m+\tilde{m})}{c^2 32\pi r} ,
\end{equation}
where $r = | \vct{x} - \tilde{\vct{x}} |$ is the distance and $\vct{v}_r$  the relative velocity of the two particles,
with the speed of light $c$ restored. Just as in \cite{Plefka:2018dpa} the post-Newtonian expansion parameter is $c^{-1}$ and the 2PN order
amounts to the order $c^{-6}$. We note the scaling properties 
\begin{gather}
 \quad (p_\mu) = (E, - \vct{p}) \sim (\Order(c^0), \Order(c^{-1})), \nonumber \\ 
\kappa \sim \Order(c^{-1})\, , \quad
\lambda \sim \Order(c^0), \quad \partial_t \sim \Order(c^{-1})\, .
\end{gather} 
The post-Newtonian expansion of the propagator becomes local in time and is expanded as
\begin{align}\label{propPN}
D(x) &= \int \frac{d^4 k}{(2\pi)^4} \frac{e^{-i k_\mu x^\mu}}{k_\mu k^\mu + i \epsilon} 
\\
&= - \int \frac{d^3 \vct{k}}{(2\pi)^3} \frac{e^{i \vct{k} \cdot \vct{x}}}{\vct{k}^2} \left[ 1 - \frac{\partial_t^2}{\vct{k}^2} + \frac{\partial_t^4}{\vct{k}^4} + \dots \right] \delta(t) .
\nonumber
\end{align}

The diagrams with the color factor $ (c\cdot {\tilde c})^3  $ have a contribution to the double-copy effective action and are extracted from the last three terms in \eqn{eq26}.
%
We further eliminate the Lagrange multipliers $\lambda_{i}, \tilde\lambda_{i}$ and worldline energies $E_{i}, \tilde E_{i}$ using their equations of motion. At 2PN, we only need the leading-order terms, and we get
\begin{align}
		L_\text{eff}^{ (c\cdot {\tilde c})^3 } = \frac{32 G^3 m^2 \tilde{m}^2}{r^3} + \frac{8 G^3 m^3 \tilde{m}}{r^3}+\frac{8 G^3 m \tilde{m}^3}{r^3},
\end{align}
where we have also replaced $\kappa$ by $\sqrt{32 \pi G}$. 

Similarly, the symmetric bulk graphs from eqs.~\eqn{eq16}--\eqn{eq19} yield
\begin{align}
\label{eq29f}
		\frac{(i\kappa)^6}{4}  \int d\hat{\tau}_{12\tilde{1}\tilde{2}}\, \left ( N_s^2 G_{12;\tilde{1}\tilde{2}} +  N_t^2 G_{1\tilde{1};2\tilde{2}} + N_u^2 G_{1\tilde{2};2\tilde{1}}
		\right ),
\end{align}
where 
\begin{align}
		N_s = \frac{1}{4} p_{1\mu} p_{2\nu} \tilde p_{1\rho} &\tilde p_{2\sigma} \Big[  V_{12}^{\mu\nu\lambda} V_{\tilde{1}\tilde{2}}^{\rho\sigma\delta} \eta_{\lambda\delta} \nonumber  \\+&\left( \eta^{\mu\rho}\eta^{\nu\sigma} - \eta^{\mu\sigma}\eta^{\nu\rho} \right) \left( \partial_1 + \partial_2 \right)^2 \Big], \label{Ns}\\
		N_t = \frac{1}{4} p_{1\mu} p_{2\nu} {\tilde p}_{1\rho} &{\tilde p}_{2\sigma} \Big[ 2\xi V_{1\tilde{1}}^{\mu\rho\lambda} V_{2\tilde{2}}^{\nu\sigma\delta} \eta_{\lambda\delta} \nonumber \\ + 2\rho &\left( \eta^{\mu\nu}\eta^{\rho\sigma} - \eta^{\mu\sigma}\eta^{\nu\rho} \right) ( \partial_{1} + {\tilde \partial_1} )^2 \Big],\label{Nt} \\
		N_u = \frac{1}{4} p_{1\mu} p_{2\nu} {\tilde p}_{1\rho} &{\tilde p}_{2\sigma} \Big[ 2(1-\xi) V_{1\tilde{2}}^{\mu\sigma\lambda} V_{2\tilde{1}}^{\nu\rho\delta} \eta_{\lambda\delta} \nonumber \\ + 2(1-\rho) &\left( \eta^{\mu\nu}\eta^{\rho\sigma} - \eta^{\mu\rho}\eta^{\nu\sigma} \right) ( \partial_1 + {\tilde \partial_2} )^2   \Big]. \label{Nu}
\end{align}
In the PN limit \eqn{Ns} reduces to $N_s = {1\over 4} E_1 E_2 \tilde E_1 \tilde E_2 (\partial_1 - \partial_2)\cdot (\tilde\partial_1 - \tilde\partial_2) + \cO(c^{-1})$ 
at leading order, i.e.~the last two terms in \eqn{Ns} do not contribute at this order. Similar reductions apply to $N_t$ and $N_u$, with an extra factor $2\xi$  in $N_t$ and $2(1-\xi)$ in $N_u$.
%
The surviving terms may indeed be made to obey the kinematical Jacobi identity at the PN level
through a suitable choice of the parameter $\xi$. One has
\begin{align}\label{Kine_Jacobi}
 N_{s}\! &-\!N_{t}\!+\!N_{u} = \frac{1}{4} E_1 E_2 \tilde E_1 \tilde E_2 \times \nonumber \\
&
\times \Bigl [ 2 (1-\xi) (\partial_1 - \tilde\partial_2)\cdot (\partial_2 - \tilde\partial_1)   \\
&\ - 2 \xi(\partial_1 - \tilde\partial_1)\cdot (\partial_2 - \tilde\partial_2)
+ (\partial_1 - \partial_2)\cdot (\tilde\partial_1 - \tilde\partial_2) \Bigr ] \nonumber \\
& + \cO(c^{-1})\, . \nonumber 
\end{align}
dropping subleading terms at $\cO(c^{-1})$. Demanding the vanishing of the above relations fixes $\xi=\frac{1}{2}$. The second introduced parameter $\rho$ for the symmetric bulk graphs remains unfixed at the 2PN
level. In fact one checks that the higher order terms in $c^{-1}$ pertaining to the PM level
\emph{cannot} be made to vanish for any choice of $\rho$. Hence, it appears necessary to
add generalized gauge transformation like terms at the 3PM level, which would, however, not affect the 2PN level considered here, {\red as is shown in appendix A}.

Hence, we take $\xi=\frac{1}{2}$ and perform the resulting two-loop integrals for the
$G_{12;34}$ functions in \eqn{eq29f} using TARCER \cite{Mertig:1998vk}
and FeynCalc \cite{Mertig:1990an,Shtabovenko:2016sxi} to find the contribution
\begin{align}
		L_\text{eff}^{ \text{sym} } = \frac{8 G^3 m^2 \tilde{m}^2}{r^3}\, .
\end{align}
Turning to the non-symmetric bulk graphs of eqs.~\eqn{eq20}--\eqn{eq21} we have
\begin{gather}
		\frac{(i\kappa)^6}{6}  \int d\hat{\tau}_{123\tilde{1}} N_{s'}^2 G_{12;3\tilde{1}}  + N_{t'}^2 G_{13; 2\tilde{1}} + N_{u'}^2 G_{1\tilde{1};23} \nonumber \\ 
		 + \text{(mirrored)} \label{eq36}
\end{gather}
where
\begin{align}
		N_{s'} = \frac{1}{4} p_{1\mu} p_{2\nu} p_{3\rho} &p_{\tilde{1}\sigma} \Big[ \alpha_1 V_{12}^{\mu\nu\lambda} V_{3\tilde{1}}^{\rho\sigma\delta} \eta_{\lambda\delta}  
		\nonumber \\
		+  \beta_1 &\left( \eta^{\mu\rho}\eta^{\nu\sigma} - \eta^{\mu\sigma}\eta^{\nu\rho} \right) \left( \partial_1 + \partial_2 \right)^2 \Big],
\end{align}
$N_{t'}$ and $N_{u'}$ have similar expressions that can be read off from \eqref{eq20} and \eqref{eq21}. The above argument for the kinematical Jacobi identity of the symmetric bulk graphs also applies to the non-symmetric graphs. This fixes $\alpha_{1,2} = 1$ and leaves $\beta_{1,2}$ arbitrary.
In the leading PN limit of \eqref{eq36} again the terms arising from the quartic vertex cancel
leaving us with
\begin{align}
	L_\text{eff}^{ \text{non-sym} }=	-\frac{4 G^3 m^3 \tilde{m}}{3 r^3}-\frac{4 G^3 m \tilde{m}^3}{3 r^3}\, ,
\end{align}
after performing the integrals.
Together with the diagrams up to 2PM, already given in \cite{Plefka:2018dpa} but now expanded to 2PN, we may assemble the double-copy prediction for the effective action.
 We further solve the equations of motion for $\lambda_{i},\tilde\lambda_{i}$ and $E_{i},\tilde E_{i}$ order by order to rewrite the action in terms of the worldline coordinates and their derivatives. 
 The resulting expression for the effective action may now be compared to the known
 result  due to Mirshekari and Will \cite{Mirshekari:2013vb}, suitably
 adjusting their parameters. 
 For a comparison it is, however, important to realize that the effective action 
 itself is gauge variant and subject to possible field redefinitions.
 Applying field redefinitions and 
 adding total derivatives, in order to match the velocity-dependent terms up to 2PN with the result of  \cite{Mirshekari:2013vb} we finally find the central result of this work
\begin{widetext}
\begin{equation}
\begin{split}
		L_{\text{eff}} = &-m-\tilde{m}+\frac{1}{2}m \textbf{v}^2+\frac{1}{2} \tilde{m} \tilde{\textbf{v}}^2 +\frac{2 G m \tilde{m}}{r}+\frac{1}{8} m \textbf{v}^4+\frac{1}{8} \tilde{m} \tilde{\textbf{v}}^4 -\frac{2 G^2 m \tilde{m}( m + \tilde{m} )}{r^2} \\
		& +\frac{G m \tilde{m}}{r} \big[\textbf{v}^2+\tilde{\textbf v}^2 -3 \textbf{v}\cdot \tilde{\textbf v} -(\textbf{n}\cdot \textbf{v}) (\textbf{n}\cdot \tilde{\textbf v}) \big] + \frac{1}{16} m \textbf{v}^6 +\frac{1}{16} \tilde{m} \tilde{\textbf{v}}^6 \\
		&+\frac{G m \tilde{m} }{r}\left(\frac{ 1}{2 }\left(\textbf{v}\cdot \tilde{\textbf{v}}\right)^2+\frac{3 }{4 }  \tilde{\textbf{v}}^4 + \frac{7}{4}\textbf{v}^2  \tilde{\textbf{v}}^2 + \frac{3 }{4 }\textbf{v}^4 -2 \textbf{v}^2  \left( \textbf{v}\cdot \tilde{\textbf{v}}\right) -2  \tilde{\textbf{v}}^2 \left( \textbf{v}\cdot \tilde{\textbf{v}} \right) \right. \\
		&\quad\left. -\frac{3 }{4 }\textbf{v}^2  \left(\textbf{n}\cdot \tilde{\textbf{v}}\right)^2 -\frac{3 }{4 } \tilde{\textbf{v}}^2 (\textbf{n}\cdot \textbf{v})^2 +\frac{3}{4} (\textbf{n}\cdot \textbf{v})^2 \left(\textbf{n}\cdot \tilde{\textbf{v}}\right)^2   +\left(  \textbf{v}\cdot \tilde{\textbf{v}} \right) \left( \textbf{n}\cdot \textbf{v}\right) \left(  \textbf{n}\cdot \tilde{\textbf{v}}\right) \right) \\
		&+G m \tilde{m}\! \left(\!\frac{3}{4} \tilde{\textbf{v}}^2   \textbf{a}\cdot \textbf{n}  \!-\! \frac{1}{4}  \left( \textbf{a}\cdot \textbf{n} \right) \left(\textbf{n}\cdot \tilde{\textbf{v}}\right)^2 \!-\! \frac{3}{2}  \left( \textbf{a}\cdot \tilde{\textbf{v}}\right) \left(  \textbf{n}\cdot \tilde{\textbf{v}}\right)   \!-\! \frac{3}{4}  \textbf{v}^2  \tilde{\textbf{a}}\cdot \textbf{n}  \!+\! \frac{1}{4}\left( \tilde{\textbf{a}}\cdot \textbf{n}\right) (\textbf{n}\cdot \textbf{v})^2  \!+\!\frac{3}{2}  \left( \textbf{n}\cdot \textbf{v}\right) \left(  \tilde{\textbf{a}}\cdot \textbf{v}\right) \!\right)\\
		&+\frac{G^2 m \tilde{m}}{r^2} \left(5 m (\textbf{n}\cdot \textbf{v})^2+m \left(\textbf{n}\cdot \tilde{\textbf{v}}\right)^2-4 m\left( \textbf{n}\cdot \textbf{v}\right) \left(  \textbf{n}\cdot \tilde{\textbf{v}}\right) +3 m \textbf{v}\cdot \tilde{\textbf{v}}-2 m \textbf{v}^2 \right. \\
		&\quad\left. +5 \tilde{m} \left(\textbf{n}\cdot \tilde{\textbf{v}}\right)^2+\tilde{m} (\textbf{n}\cdot \textbf{v})^2-4  \tilde{m} \left( \textbf{n}\cdot \textbf{v} \right) \left( \textbf{n}\cdot \tilde{\textbf{v}}\right) +3 \tilde{m} \textbf{v}\cdot \tilde{\textbf{v}}-2  \tilde{m} \tilde{\textbf{v}}^2 \right)\\
		&+G^3 \left(\frac{14 m^3 \tilde{m}}{3 r^3}+\frac{26 m^2 \tilde{m}^2}{r^3}+\frac{14 m \tilde{m}^3}{3 r^3}\right)\, .
		\label{finalresult}
\end{split}
\end{equation}
\end{widetext} 
Indeed, the 2PN static terms are seen to differ from the result $L^{MW}_{\text{eff}}$
of Mirshekari and Will \cite{Mirshekari:2013vb} by 
$\Delta L_{\text{eff}}:=
L_{\text{eff}}-L^{MW}_{\text{eff}}$ \footnote{Their parameters are adjusted as $\alpha=1$, $\bar{\gamma}=-1$, $\bar{\beta}_{1,2}=0$, $\bar{\delta}_{1,2}=\frac{1}{4}$, $\bar{\chi}_{1,2}=0$. Also note that converting to our convention needs to take $G^{MW} \rightarrow 2 G$ } 
\begin{align}
\Delta L_{\text{eff}}=	\frac{2 G^3 m \tilde{m} (m+\tilde{m})^2}{r^3}
\end{align}

This disagreement cannot be removed by either field redefinition or adding total derivatives. 
It is conceivable that an adjustment of the gauge fixing condition in the original Yang-Mills
theory (at higher orders in $g$), which would modify the double copy result, could remedy this
disagreement. The important point here is that the effective action is a gauge \emph{variant} 
quantity. However, such a remedy would question the usefulness of the double-copy
procedure.


We have also performed an independent check of the validity of 
the effective potential of \cite{Mirshekari:2013vb} by performing a probe limit.
Here one takes $m\ll \tilde m$ and compares to the potential experienced by a test
particle of mass $m$ in the JNW solution of dilaton-gravity \cite{Janis:1968zz} which
is the relevant black hole solution in this setup as shown in \cite{Luna:2016hge}. We found agreement with the static terms 
linear in $m$ of \cite{Mirshekari:2013vb} and disagreement with our double copy results
in \eqn{finalresult}.
On top we also performed a full perturbative computation in dilaton-gravity at 2PN order
of the static terms again reproducing \cite{Mirshekari:2013vb}. In summary, 
the breakdown of the
double copy procedure for the effective potential is therefore on firm grounds.

\acknowledgements
We thank Lance Dixon and Wadim Wormsbecher for discussions and comments.
This project has received funding from the European Union's Horizon 2020 research and innovation programme under the Marie Sklodowska-Curie grant agreement No. 764850,
and the German Research Foundation through grant PL457/3-1.

\appendix

{  \red
\section{Implementing the kinematic Jacobi identity at the 3PM level}

In section V we showed that the kinematic numerators of the symmetric bulk graphs
as well as the non-symmetric ones obey the kinematic Jacobi identities only 
at the leading order in the PN limit, $N_{s}-N_{t}+N_{u}
=\cO(c^{-1})$ . We shall now construct an explicit representation of these bulk graphs that obeys the Jacobi identity by a generalized gauge transformation \cite{Bern:2010yg} and prove that
it will not affect the 2PN answer \eqn{finalresult} above. 

After a Fourier transform of the bulk degrees of freedom the symmetric graphs 
in Yang-Mills theory eqs.~\eqn{eq16}-\eqn{eq19} take the form
\begin{align}I^{\text{sym}}=&
\int d^{16}\hat q \int d\hat{\tau}_{12\tilde{1}\tilde{2}} \Bigl [  
\frac{c_{s}\, N_{s}}{s}+ \frac{c_{t}\, N_{t}}{t}+ \frac{c_{u}\, N_{u}}{u}
\Bigr ]\,\nn\\ &\times \frac{1}{q_{1}^{2}q_{2}^{2}q_{\tilde{1}}^{2}q_{\tilde{2}}^{2}}\, , 
\end{align}
with $s=(q_{1}+q_{2})^{2}$, $t=(q_{1}+q_{\tilde{1}})^{2}$, $u=(q_{1}+q_{\tilde{2}})^{2}$,
\be
d^{16}\hat q = \prod_{i=1,2,\tilde{1},\tilde{2}}d^{4}q_{i}\, 
e^{i\, q_{i}\cdot x_{i}} 
\ee
and
\begin{align}
c_{s}=&f^{abe} f^{cde} c_{1a} c_{2b} \tilde c_{1c} \tilde c_{2d}\, , \quad
c_{t}=&f^{ace} f^{bde} c_{1a} c_{2b} \tilde c_{1c} \tilde c_{2d}\nn\\
c_{u}=&f^{abe} f^{cde} c_{1a} c_{2b} \tilde c_{1c} \tilde c_{2d} \, . 
\end{align}
Moreover, the kinematical numerators $N_{s,t,u}$ in the above are given by the expressions  \eqn{Ns}, \eqn{Nt} 
and \eqn{Nu} with the
substitution $\partial_{i}\to i\, q_{i}$. Clearly, the color factors  obey the 
Jacobi identity
\be \label{jack}
c_{s}-c_{t}+c_{u}=0\, .
\ee
However, the kinematical numerators fail to fulfill this identity
\be
N_{s}-N_{t}+N_{u}=\Delta(p_{i},q_{i})\, ,
\ee
with a lengthy expression $\Delta(p_{i},q_{i})$ following from \eqn{Ns}, \eqn{Nt} 
and \eqn{Nu}. We now introduce the generalized gauge transformations 
\be \tilde{N}_{a}= N_{a}
+\Delta_{a}\, , \qquad a=s,t,u\ee which need to obey two conditions: The first one is
\be
 \Delta_{s}-\Delta_{t}+\Delta_{u}= -\Delta \label{(1)} 
 \ee
 securing the Jacobi identity for the $\tilde{N}_{a}$. The second condition
demands to leave $I^{\text{sym}}$ invariant
 \begin{align}
  0=\int &d^{16}\hat q \int d\hat{\tau}_{12\tilde{1}\tilde{2}}\Bigl ( c_{s}\Bigl [  
\frac{\Delta_{s}}{s}+ \frac{\Delta_{t}}{t} \Bigr ] 
+ c_{u}\Bigl [ \frac{\Delta_{u}}{u}+ \frac{\Delta_{t}}{t} \Bigr ] \Bigr ) \nn\\ &
\times
\frac{1}{q_{1}^{2}q_{2}^{2}q_{\tilde{1}}^{2}q_{\tilde{2}}^{2}}\, , \label{(1)}
\end{align}
having replaced $c_{t}=c_{s}+c_{u}$ by \eqn{jack}. Demanding the vanishing
of this at the integrand level yields the simple solutions
\begin{align}
\Delta_{s}&= -\frac{s\, \Delta}{s+t+u}\, , 
\qquad
\Delta_{t}=  \frac{t\, \Delta}{s+t+u}\, , \nn\\
\Delta_{u}&= -\frac{u\, \Delta}{s+t+u}\, . \label{delsol}
\end{align}
Of course the vanishing of \eqn{(1)} would in principle also allow for total derivative contributions
in the integrands. However,  \emph{all} solutions will lead to identical double copy results
so it is sufficient to stick with the simple ones of \eqn{delsol}.
Now as $\Delta= \mathcal{O}(c^{-1})$ -- as was shown in \eqn{Kine_Jacobi} for the choice $\xi=\frac{1}{2}$ -- and the fact that $s,t,u$ are independent of $c$ or $\kappa$
we immediately conclude that the kinematical Jacobi relation respecting numerators
$\tilde{N}_{a}$ differ from the original $N_{a}$ only by terms of order $\mathcal{O}(c^{-1})$.
This implies in turn that the contributions from the double copied symmetric graphs to
the putative effective action of dilaton gravity obtained from
the $\tilde{N}_{a}$ does not differ from the one constructed by the $N_{a}$ computed in \eqn{finalresult}
at the 2PN level
\begin{align}
 & \int d\hat{\tau}_{12\tilde{1}\tilde{2}} \, \Bigl ( \tilde{N}_s^2 G_{12;\tilde{1}\tilde{2}} +  \tilde{N}_t^2 G_{1\tilde{1};2\tilde{2}} + 
		\tilde{N}_u^2 G_{1\tilde{2};2\tilde{1}} \Bigr )	\nn\\
		&=  \int d\hat{\tau}_{12\tilde{1}\tilde{2}} \, \Bigl ( N_s^2 G_{12;\tilde{1}\tilde{2}} +  N_t^2 G_{1\tilde{1};2\tilde{2}} + 
		N_u^2 G_{1\tilde{2};2\tilde{1}} \Bigr )\nn\\ & \quad +\mathcal{O}(c^{-1})\, .
\end{align}
The analogous argument goes through for the non-symmetric bulk graphs. 

In summary, we have thus
shown that implementing the kinematical Jacobi identity at the 3PM level does not
affect the 2PN results reported in the main text.

}

\end{fmffile}

\bibliographystyle{apsrev4-1}
\bibliography{inspire}
\newpage

\onecolumngrid


\end{document}